\documentclass[a4paper,11pt,twoside,openright,onecolumn]{article}
\usepackage{graphicx}
\usepackage{pdflscape}
\usepackage{natbib}  
\usepackage{authblk}

 \def\.epss{Ap\&SS}%
 \def\a.eps{A\&AS}%
 \def\.epspr{Astrophys.~Space~Phys.~Res.}%
\renewcommand{\rmdefault}{cmss} 
\fontfamily{\rmdefault}
\selectfont

\begin{document}

\title{INPOP new release: INPOP10e}

 \author[1,2]{A. Fienga\thanks{agnes.fienga@obs-besancon.fr}}
 \author[1]{H. Manche} 
 \author[1]{J. Laskar}
 \author[1]{M. Gastineau}
 \author[2]{A. Verma}
\affil[1]{Astronomie et Syst\`emes Dynamiques, IMCCE-CNRS UMR8028, 77 Av. Denfert-Rochereau, 75014 Paris, France}
\affil[2]{Institut UTINAM-CNRS 6213, Universit\'e de Franche-Comt\'e, 41bis Av. de l'Observatoire, 25000 Besan\c con, France}



 \date{}


 
  \maketitle


\begin{abstract}
The INPOP ephemerides have known several improvements and evolutions since the first INPOP06 release (Fienga et al. 2008) in 2008. In 2010, anticipating the IAU 2012 resolutions, adjustement of the gravitational solar mass with a fixed astronomical unit (AU) has been for the first time implemented in INPOP10a (Fienga et al. 2011)  together with improvements in the asteroid mass determinations. With the latest INPOP10e version, such advancements have been enhanced and studies about solar corona have also been investigated (Verma et al. 2012). The use of planetary ephemerides for several physical applications are presented here from electronic densities of solar slow and fast winds to asteroid mass determinations and tests of general relativity operated with INPOP10a. Perspectives will also be drawn especially related to the analysis of the Messenger spacecraft data for the planetary orbits and future computation of the time variations of the gravitational mass of the sun. 
\end{abstract}



\section{Introduction}

Since 2006,  INPOP (Integration Numerique Planetaire de l'Observatoire de Paris) has  become an international reference for space navigation (to be used for the GAIA mission navigation and the analysis of the GAIA observations) and for scientific research in  dynamics of the solar system objects and in fundamental physics. 
A first version of INPOP, INPOP06, was published in 2008 (Fienga et al. 2008). This version is very close to the reference ephemerides of JPL in its dynamic model and in its fit procedure.  With MEX and VEX tracking data provided by ESA, lunar laser ranging observations and the development of new planetary and moon ephemeris models and new adjustment methods, INPOP08 (Fienga et al. 2009) and INPOP10a (Fienga et al. 2011) were constructed. These versions of INPOP have established INPOP at the forefront of global planetary ephemerides because its precision in terms of extrapolation to the position of planets is equivalent to the JPL ephemerides. Its dynamic model follows the recommendations of the International Astronomical Union (IAU) in terms of i) compatibility between time scales (TT, TDB), ii) metric in the relativistic equations of motion (consistency in the computation of the position of the barycenter of the solar system) and iii) in the fit of the sun gravitational mass with a fixed AU.

INPOP provides to the user, positions and velocities of the planets, the moon, the rotation angles of the earth and the moon as well as TT-TDB chebychev polynomials at $\textrm{http://www.imcce.fr/inpop}$.
INPOP10a was the first planetary ephemerides in the world built up with a direct estimation of the gravitational mass of the sun with a fixed astronomical unit instead of the traditional adjustment of the AU scale factor.
With INPOP10a, we have demonstrated the feasibility of such determination helping the IAU of taking the decision of fixing the astronomical unit (see resolution B2 of the 35th IAU general assembly, 2012).

The INPOP01e is the latest INPOP version developed for the Gaia mission final release and available for users. 
Compared to INPOP10a, new sophisticated procedures related to the asteroid mass determinations have been implemented: bounded value least squares have been associated with a-priori sigma estimators (Kuchynka 2010, Fienga et al. 2011)
and solar plasma corrections (Verma et al. 2012).
Very recent Uranus observations provided by (Viera Martins and Camaro 2012) have been added as well as positions of Pluto deduced from HST (Tholen et al. 2008). 
For the LLR fit, additionnal observations are available from Cerga, MLRS2, Matera ($\textrm{ftp:/cddis.gsfc.nasa.gov/slr/data/npt/moon}$) and Apollo ($\textrm{http://physics.ucsd.edu/~tmurphy/apollo}$).

Adjustment of the gravitational mass of the sun is performed as recommended by the IAU resolution B2 as well as the sun oblateness (J$_{2}$), the ratio between the mass of the earth and the mass of the moon (EMRAT) and the mass of the Earth-Moon barycenter. Estimated values are presented on Table \ref{paramfita}.  

Masses of the planets have been as well updated to the IAU best estimated values (Luzum et al. 2012). 

\begin{table}
\caption{Values of parameters obtained in the fit of INPOP10e and INPOP10a to observations.}
\begin{center}
\begin{tabular}{l c c c }
\hline
&  INPOP10e & INPOP06 & DE423 \\
&    $\pm$ 1$\sigma$ & $\pm$ 1$\sigma$ & $\pm$ 1$\sigma$ \\
\hline
(EMRAT-81.3000)$\times$ 10$^{-4}$ &  (5.700 $\pm$ 0.020) & 5.6 & (5.694 $\pm$ 0.015) \\

J$_{2}$$^{\odot}$ $\times$ 10$^{-7}$ & (1.80 $\pm$ 0.25) & (1.95 $\pm$ 0.5)  & 1.80  \\
\hline
GM$_{\odot}$ - 132712440000 [km$^{3}.$ s$^{-2}$]&  (50.16 $\pm$ 1.3) & 17.987 & 40.944 \\
AU - 1.49597870700 $\times$ 10$^{11}$ [m] & 9.0 & 9.0 & (-0.3738 $\pm$ 3 ) \\
\hline
[M$_{\odot}$ / M$_{\textrm{EMB}}$] - 328900    & 0.55223 $\pm$ 0.004  & 0.56140  & 0.55915 $\pm$ NC \\
 \hline
 \end{tabular}
\end{center}
\label{paramfita}
\end{table}

Thanks to the added solar corrections and to the improvement in the fit procedure, 152 asteroid masses have been estimated (see section \ref{secapp}). 
Comparisons to other planetary ephemerides, postfit and extrapolated residuals are discussed in section \ref{uncert}.
  
\section{Estimation of uncertainties}
\label{uncert}

\subsection{Comparisons to other planetary ephemerides}

In order to better estimate the INPOP10e uncertainties, comparisons are made  between INPOP10e, INPOP10a and the JPL DE423 (Folkner 2010) in spherical coordinates (table \ref{raderho}) for the planets relative to the earth and in cartesian coordinates (table \ref{xyz}) for the earth relative to the solar system barycenter in the ICRF (also called BCRS) over a period of 20 years before and after J2000.
With these figures, differences in the dynamic model, fitting procedures and data sample can be impacted on planetary positions and velocities for an interval of time corresponding to the most accurate data sets.

The DE423 ephemerides have been fitted on a data set similar to the INPOP10e one. Fitting procedures differ with less asteroid masses adjusted in DE423 (63) and smoother behavior in the Mars residuals during the fitted period (see table \ref{comparomc}). 
INPOP10e  differs from INPOP10a by new corrections in the Messenger data, new implementation in the fit of the asteroid masses and in the correction of the solar plasma, and the use of very recent observations of Uranus (Viera Martins and Camaro 2012) inducing modifications in the weighting schema of the adjustment. Differences between INPOP10e, DE423 and INPOP10a can be seen as good estimations of the state-of-art uncertainties of planetary ephemerides.

As expected, the uncertainties of the positions of inner planets are quite smaller than those obtained for the outer planets. 
This can easily be explained by the use in the ephemeris construction of high accurate data deduced from the tracking of inner planet orbiters during the past 40 years. 

For Jupiter, the uncertainty in geocentric distance is about 1 km but the angular differences are not quite similar from one ephemeris to another: from 10 mas with INPOP10a to less than 1 mas with DE423. 

Due to these important variations and to the expected lack of accurate Jupiter observations in the near future, the accuracy of the Jupiter orbit is very likely to degrade in the coming years.
For Saturn, the ephemerides give more consistent results reflecting the important role of the Cassini observations in the Saturn orbit determination.
For Uranus, Neptune and Pluto, the important differences illustrate the lack of accurate estimations of distances and angular positions for these objects.

Differences in the earth BCRS positions and velocities obtained for several planetary ephemerides (see table \ref{xyz}) are about  1 kilometer in positions and smaller than 0.1 mm.s$^{-1}$ in velocities. Comparisons between DE423 and DE421 (Folkner et al. 2008) which differ mainly by the data sample are equivalent to those obtained with the two consecutive INPOP versions (INPOP10e and INPOP10d (Verma et al. 2012)). In the case of INPOP10e and INPOP10a, these figures can be explained up to 85 $\%$ by differences in the estimation of the gravitational mass of the sun.

\subsection{Comparisons to observations, extrapolation and link to the ICRF}
\label{omcsec}

\subsubsection{Planetary observations}
The INPOP10e observational sample has 3 times more data than the INPOP06 one (the first INPOP release) which ended in 2005.45. The statistical distribution of the supplementary data sets is not uniform and is mostly constituted with MEX and VEX observations (60 $\%$). However, the two flyby points of Uranus and Neptune and the five flybys of Jupiter are of crucial importance for the accuracy of these orbits. The three positions of Mercury deduced from the Messenger flybys play also an important role for the Mercury orbit determination even if their distribution in time was very limited (less than 2 years). 

On tables \ref{omctab0} and \ref{omctab1} are given some examples of postfit and extrapolated residuals obtained with INPOP10e and other ephemerides. For Mars, INPOP10e faces an improvement of the extrapolated residuals compared to INPOP10a and obtains the same level of accuracy as the JPL DE423.
The Saturn residuals presented  in table \ref{omctab0} are good examples of the improvement of the outer planet orbits obtained with INPOP10e compared to the previous INPOP versions. In particular, a reduction of a factor more than 10 is obtained in Cassini range residuals. This improvement is also confirmed with the Uranus and Neptune flyby residuals as one can see on table \ref{omctab1}. By providing measured distances between the earth and the outer planets, the flyby data brought new informations to the fit when only optical observations were used in the INPOP06 and INPOP08 adjustments. 
As a result, one can notice on table \ref{omctab1} the satisfactory INPOP06 residuals obtained for the outer planet flyby residuals in right ascension and declination (at the level of the accuracy of the optical data used in the INPOP06 fit) but the very poor estimations in distances. 


For Jupiter, the expected accuracy of the ephemerides will not be better than the postfit residuals obtained by comparison to flyby positions which reach up about 2 kilometers (see table \ref{omctab0}).  Unfortunately, no direct accurate observation of Jupiter (such as radio or VLBI tracking of a spacecraft in its vicinity) are planned in the near future in order to maintain the constraints over the Jupiter orbit. Calibration of possible Jupiter orbit degradation would only be partially possible through indirect constraints from Cassini Solstice mission, Dawn, Messenger, present and future Mars orbiters. 
However, contrary to Jupiter, new Saturn positions would be obtained during the Cassini Solstice mission through 2017 and would then be helpful for constraining the Saturn orbit in the coming years.
 
\begin{figure}
\includegraphics[scale=0.3]{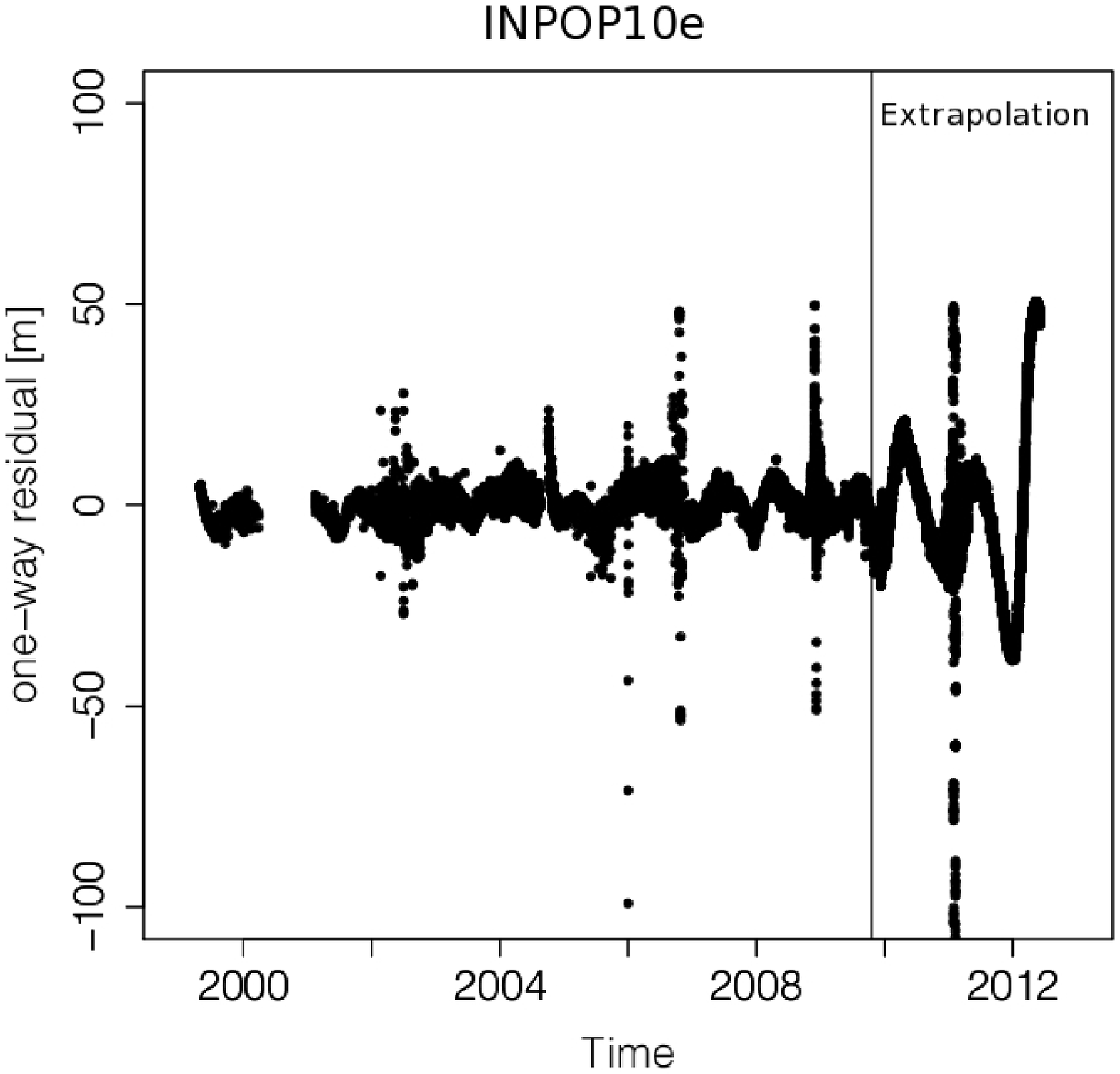}\includegraphics[scale=0.3]{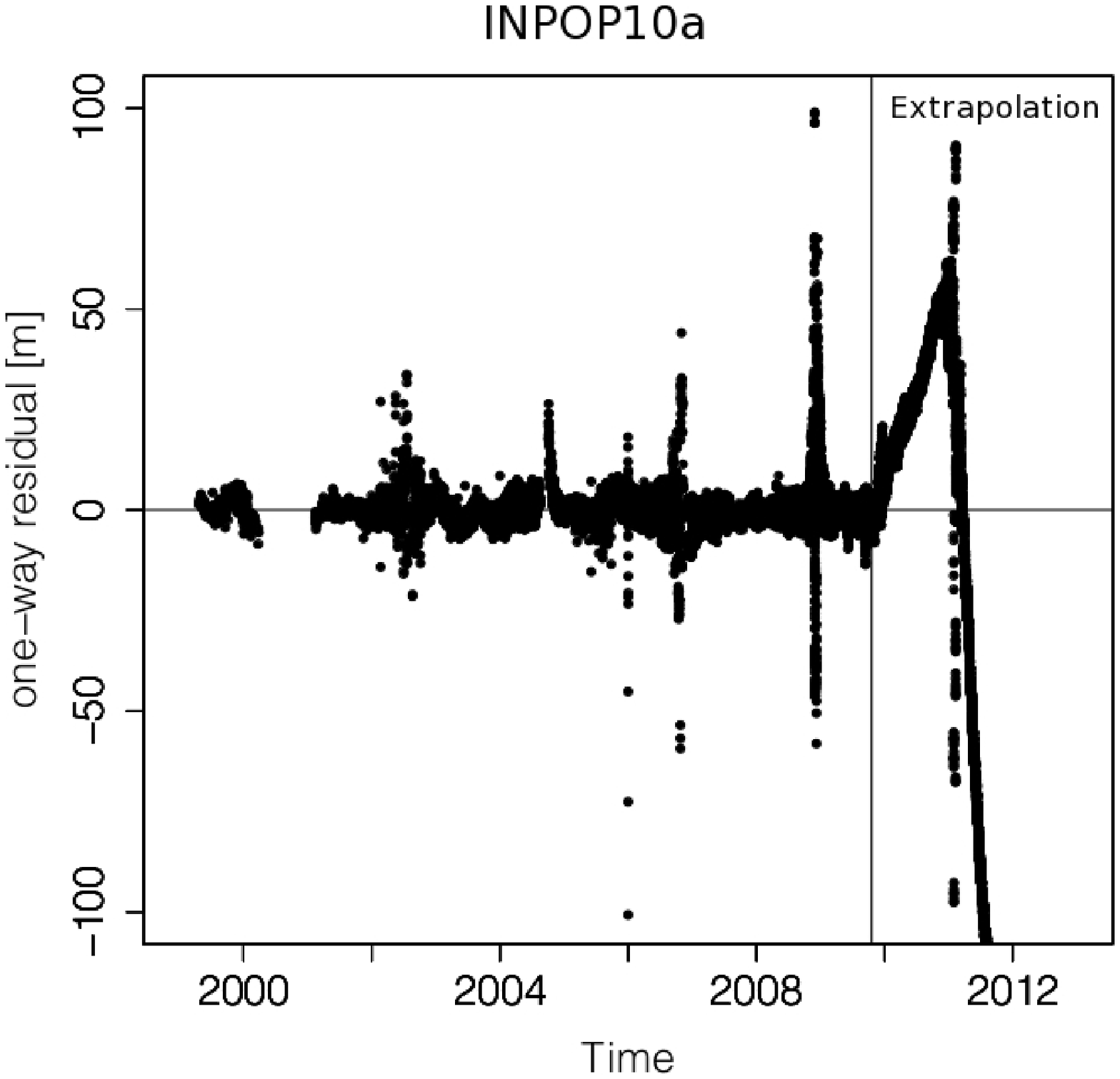}\includegraphics[scale=0.3]{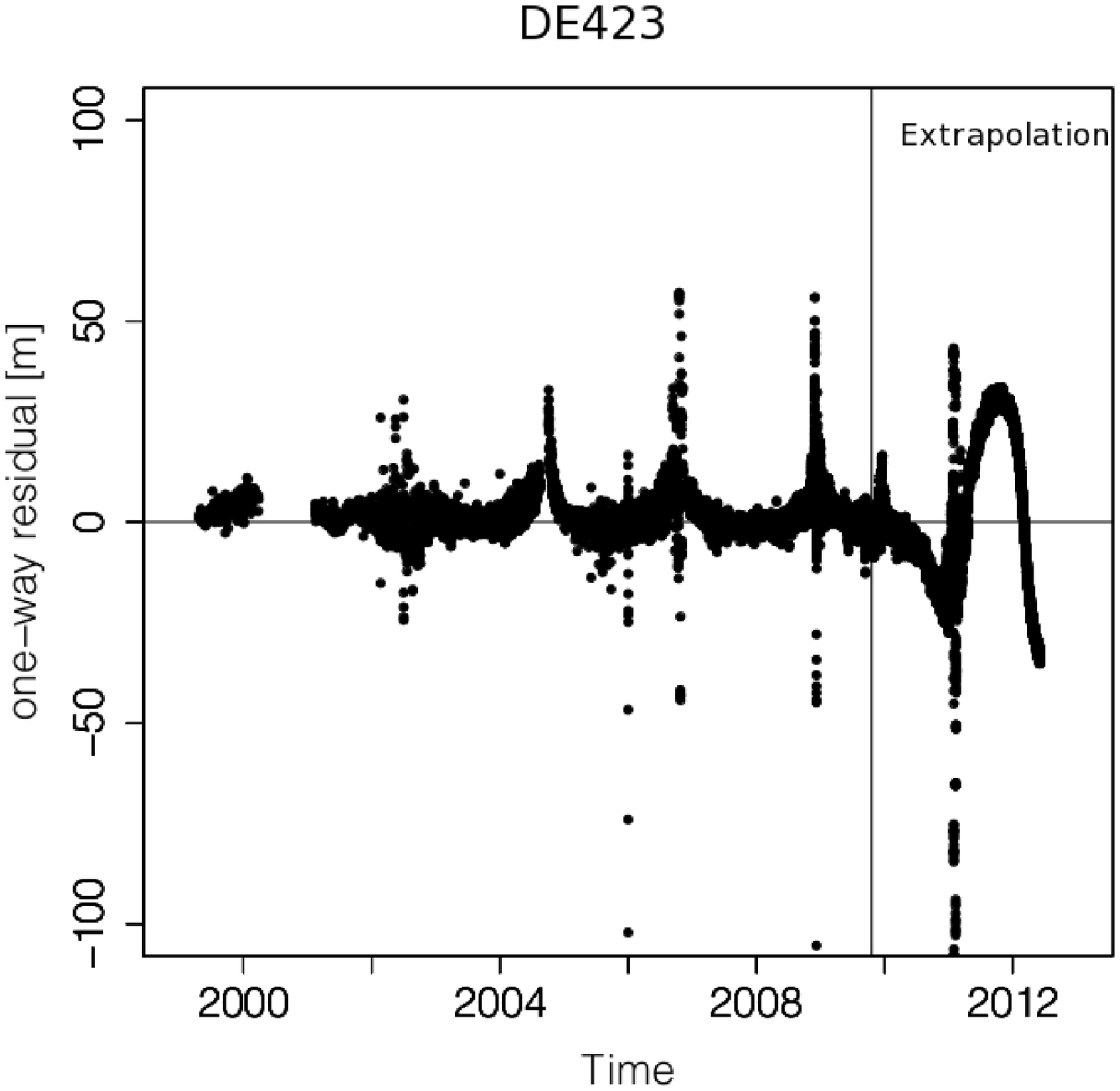}\\
\caption{Some examples of postfit and extrapolated MEX one-way range residuals in meters obtained with INPOP10e, INPOP10a, INPOP06 and DE423.}
\label{comparomc}
\end{figure}

For the inner planets, the orbits are very well constrained thanks to spacecraft tracking data of Mars orbiters, VEX and Messenger missions. However, we note a rapid degradation of the Mars orbit accuracy as estimated by comparison between planetary ephemerides and observed MEX distances not included in the fit of the ephemerides. Such comparisons are called extrapolation in the figure \ref{comparomc} and in table \ref{omctab1}. The differences  between estimated distances and the observed one reach up to 30 meters after 32 months and are mainly due to un-modeled perturbations of main-belt asteroids.

Even if not seen as a major planet anymore, Pluto orbit is also included in the INPOP planetary ephemerides. For our latest version, we work on the improvement of the Pluto orbit in including stellar occultations (as in INPOP10a) but also positions of the Pluto-Charon barycentric system obtained in 2008 with HST by (Tholen et al. 2008). In the opposite of DE423, INPOP10e shows un-biased residuals in right ascension and declination as one can see on table \ref{omctab1}. 

The tie between INPOP ephemerides and the ICRF (McCarthy and Petit, 2003) is maintained by the use of VLBI differential observations of spacecraft relative to ICRF sources. Such methods give milliarcsecond (mas) positions of a spacecraft orbiting a planet directly in the ICRF. Combining such VLBI observations with spacecraft navigation, positions of planets can be deduced relatively to the ICRF sources. The link between modern planetary ephemerides and the ICRF is then obtained at the accuracy of the VLBI localization of the space missions. Based on the most recent Mars, VEX and Cassini VLBI observations, the link between the INPOP10e reference frame and the ICRF is maintained with an accuracy of about 1 mas for the last 10 years. 

\begin{table}
\caption{Maximum differences between INPOP10e, INPOP10a and DE423 from 1980 to 2020 in spherical geocentric coordinates and distances.}
\begin{center}
\begin{tabular}{c | c c c | c c c}
\hline
Geocentric & \multicolumn{3}{c}{INPOP10e - INPOP10a} & \multicolumn{3}{c}{INPOP10e - DE423} \\
Differences & \multicolumn{3}{c}{1980-2020} & \multicolumn{3}{c}{1980-2020} \\
& $\alpha$ & $\delta$ & $\rho$ & $\alpha$ & $\delta$ & $\rho$  \\
& mas & mas & km & mas & mas & km \\
Mercury & 1.4 & 3.1 & 0.6 & 1.58 & 1.7 & 0.65   \\
Venus & 0.27 & 0.43 & 0.021 & 0.85 & 0.42 & 0.045 \\
Mars & 1.26 & 0.37 & 0.185 & 2.1 & 0.62 & 0.47 \\
Jupiter & 4.13 & 9.94 & 0.88 & 0.81 & 0.74 & 1.11  \\
Saturn & 0.54 & 0.52 & 0.51 & 0.82 & 0.53 & 1.82 \\
Uranus & 226.9 & 120.2 & 1370 & 98.1 & 38.9 & 359.73 \\
Neptune & 12.6 & 6.5 & 1081 & 51.0 & 91.3 & 2054.8  \\
Pluton & 25.53 & 154.8 & 3447.1 & 703.2 & 152.7 & 37578.6  \\
\hline
\end{tabular}
\label{raderho}
\end{center}
\caption{Maximum differences between INPOP10e and other planetary ephemerides from 1980 to 2020 in cartesian coordinates of the earth in the BCRS.}
\begin{center}
\begin{tabular}{c | c c}
\hline
Earth Barycentric &  XYZ & VxVyVz\\
Differences &   & \\
& km  & mm.s$^{-1}$ \\
\hline
INPOP10e - INPOP10a & -1.0 & 0.050 \\
INPOP10e - DE423 & 0.84 &  0.113 \\
DE423-DE421 & 0.37 & 0.070 \\
INPOP10e-INPOP10d & 0.34  &  0.050 \\
\hline
\end{tabular}
\label{xyz}
\end{center}
\end{table}


\subsubsection{LLR observations}
Concerning Lunar Laser Ranging, INPOP10e takes into account observations from 1969 to 2012. Compared to INPOP10a, 211 observations from CERGA, 255 from MLRS2, 300 from Apollo and 33 from Matera have beed added in the fit. The number of parameters fitted is now 65, including the same 59 that were already fitted in INPOP10a. The 6 other parameters are for the positions of Matera station, and the positions of the Lunakhod 1 reflector. This latter was lost for about 40 years, and has been retrieved in 2010 by Apollo (Murphy 2010). The values of all fitted parameters are given in Tables \ref{Tab_valeurs_parametres_LLR_dyn_I10e}, \ref{Tab_valeurs_parametres_LLR_I10e_reflecteurs} and \ref{Tab_valeurs_parametres_LLR_I10e_stations}. 
Their formal errors (1$\sigma$) come from the covariance matrix of the least square fit and can be much smaller than the physical uncertainties. A test has been performed with the sum of gravitational constants of the Earth and the Moon (called $GM_{EMB}$), showing that modifying its value of about 16 times the formal $\sigma$ leads, after the adjustment of all the other 64 parameters, to the increase of the $\chi^2$ of 1\% (and of 5\% if $GM_{EMB}$ is modified of 32 times $\sigma$).

On Table \ref{Tab_residus_LLR_I10e} and figure \ref{FigResidusLLR10e} are given the Lunar Laser Ranging residuals.

For Apollo, the residuals obtained with INPOP10e are slightly degraded (3 millimeters) compared to the ones obtained with INPOP10a. 
This degradation can have several possible explanations. However, we stress the fact that the Apollo station is not included in the International Terrestrial Reference Frame (ITRF): the velocity vector of the Apollo station coordinates (modeling the tectonic plate motion) is thus unknown. In INPOP10e, its value has been fixed to the same as the closest ITRF station (White Sands, far away from 65 kilometers), but the real value for Apollo could be slightly different. 


For the first period of MLRS2 (before 1996), the degradation of 5 millimeters noticed for INPOP10e compared to INPOP10a is mostly induced by the differences in the Earth Orientation Parameters (EOP given by the C04 series) between ITRF2008 and ITRF2005. These differences can reach up to 5 mas for Celestial Intermediate Pole corrections dX and dY. Despite the fit of the station coordinates and the update of the velocity vectors from ITRF 2005 to ITRF 2008, the increase of residuals can not be eliminated. On the same period,  observations from Cerga are also available. But these latest are not sensitive to this change of EOPs, certainly because the standard deviation of the residuals (6.3 centimeters) is higher than for MLRS2 (4.7 centimeters).

For the second period of MLRS2 (after 1996), the observations are very noisy: more than 21\% of data are eliminated according to a 3\_$\sigma$ criterion (less than 6\% for the other stations, except Matera).
If all of them had been kept in the fit, standard deviation would have reached the value of 5 meters, whereas observations from Cerga on the same period have a standard deviation of about 4 centimeters (with less than 2\% of them eliminated).
Furthermore, the standard deviation of MLRS2 observations after 2008 (newly added in INPOP10e) is significantly higher by almost a factor 2 than the one before 2008 (already taken into account in INPOP10a). 

In conclusion, the degradation of the residuals cannot be clearly explained at this point of the investigations but lacks in the dynamical and/or reduction model could create such behaviors.
Meanwhile, (Murphy and Adelberger 2011) have noticed that for some nights, when several reflectors have been observed, there are offsets in residuals depending on the reflector (see figures \ref{FigResidusLLR10e2Nuits}).  This could also be explained by a mis-modeling in the orientation of the Moon, and work is in progress in order to improve the dynamical model of its rotation by adding a lunar core.

\section{Applications}
\label{secapp}

\subsection{Solar physics}

\begin{figure}
 \includegraphics[scale=0.4]{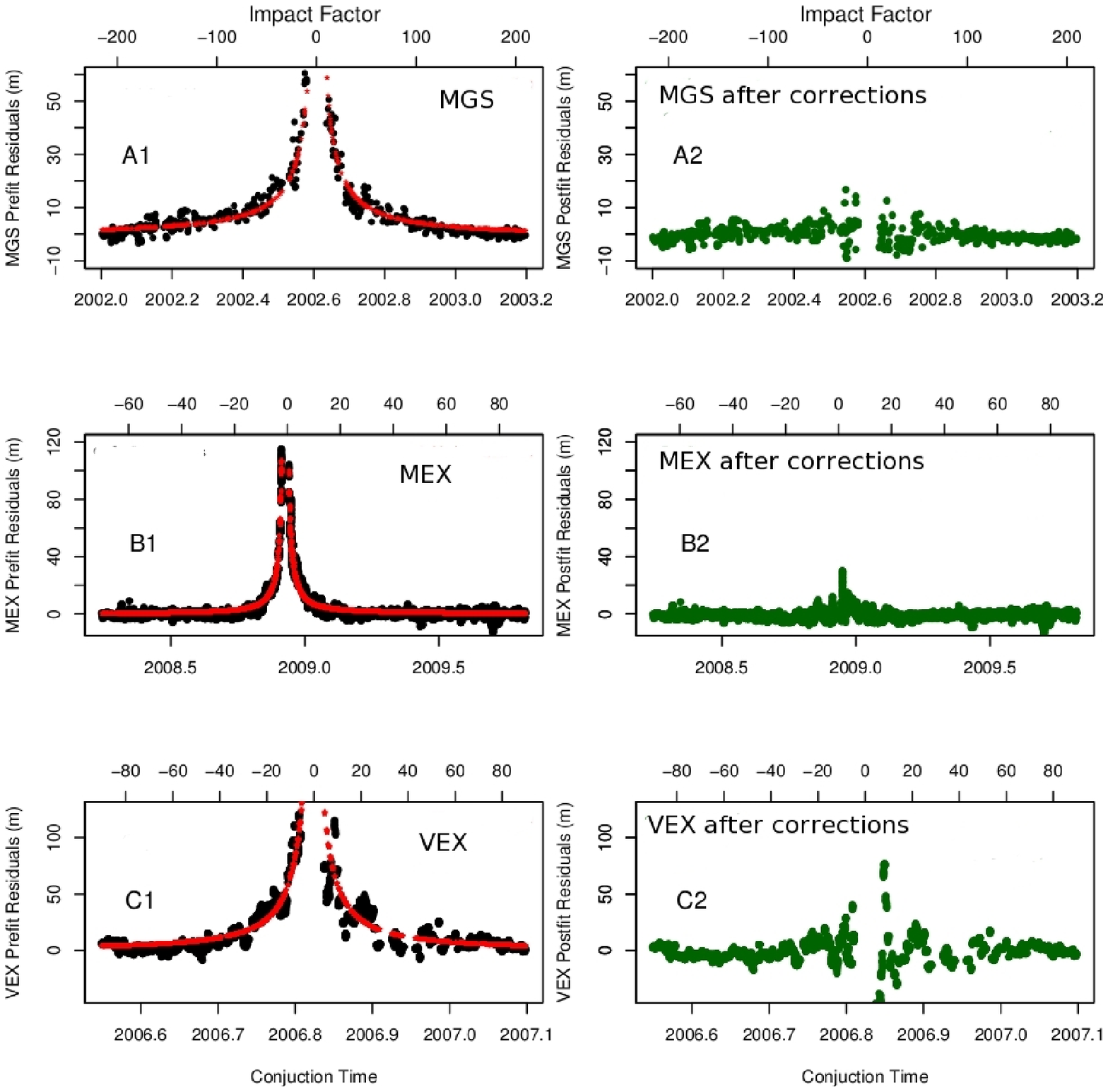}\includegraphics[scale=0.34]{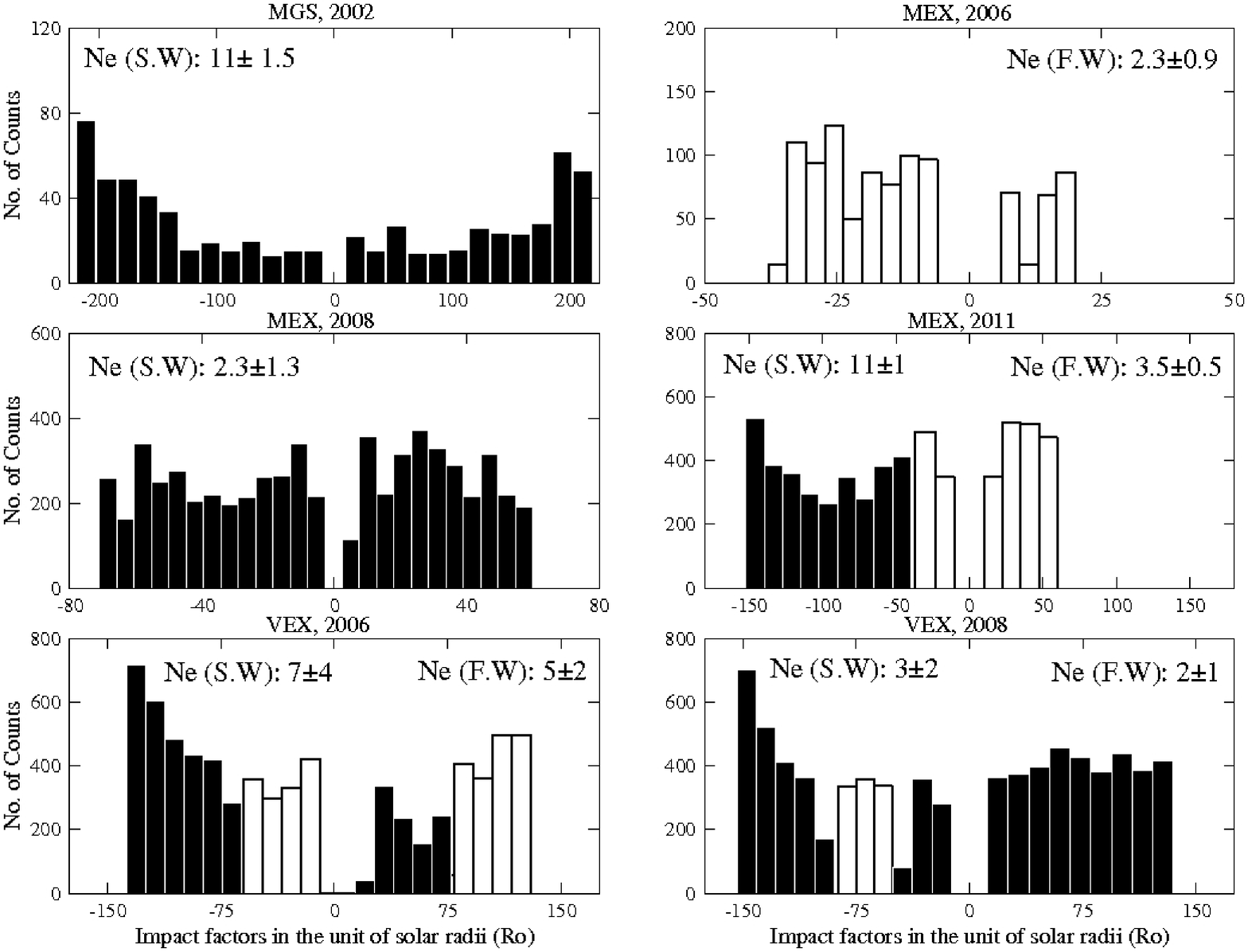} 
 \caption{Solar physics applications: a) Corrections of solar plasma applied to MGS, MEX and VEX range measurements. b) Histograms of the distributions of MGS, MEX and VEX data during solar conjunctions in slow (black) and fast (white) wind regions. Are also given the values of the obtained electronic density ($N_e$) at 1 AU in electrons.cm$^{-3}$.}
 \label{solphy}
\end{figure}

As one can see on the left-hand side chart of the figure \ref{solphy}, range observations of MGS, MEX and VEX spacecraft were highly affected by solar plasma perturbations during solar conjunctions, but also before and after these critical periods. In the opposite side of the spectrum, solar physicists are interested in characterizing electronic densities of two specific area on the sun surface: the regions in which dominates a slow wind (mainly following the magnetic neutral line) and the regions (higher in solar latitudes) corresponding to fast winds (Schwenn and Marsch 1990, 1991). By analyzing the path of the radiometric signal from the spacecraft to the earth, it is possible to estimate such electronic densities for the two regions during the ingress and the egress parts of the signal and for different phases of the solar activity (Verma et al. 2012). On the right-hand chart of the figure \ref{solphy}, the distributions of MGS, MEX and VEX analyzed data in slow and fast wind regions are plotted and the obtained electronic densities for the two regions are also given. 
After the estimation of the electronic densities, solar plasma corrections were applied to the radiometric signal from the Mars and Venus orbiters, as one can see on the left-hand side chart of the figure \ref{solphy}. Such corrections allow to re-introduce in the INPOP fit 8$\%$ of supplementary data, previously rejected, and then to improve the extrapolation capabilities of INPOP10e and the asteroid mass determinations (Verma et al. 2012). 

\subsection{Asteroid masses}

\begin{figure}
\begin{center}
\includegraphics[scale=0.5]{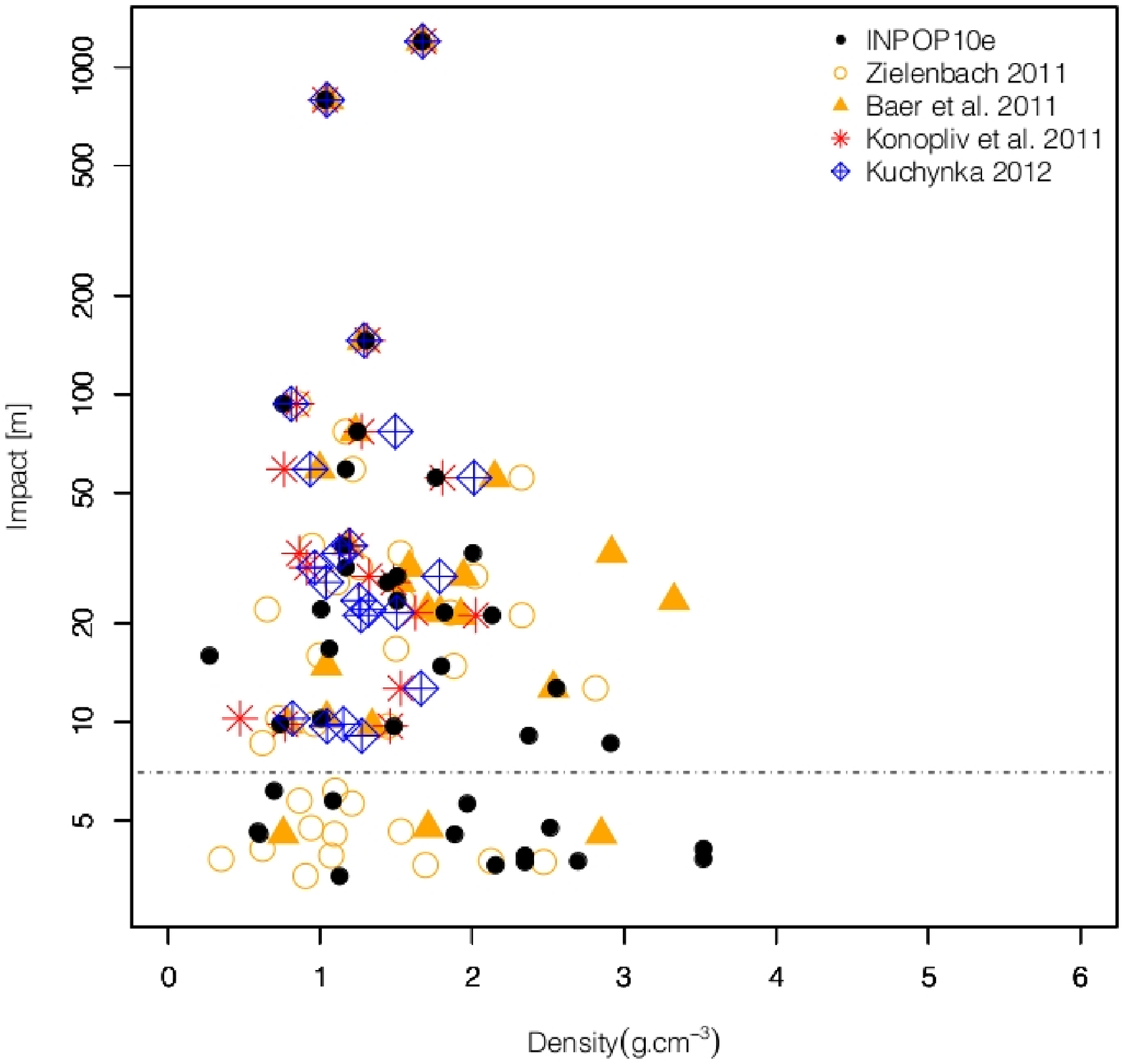}\includegraphics[scale=0.5]{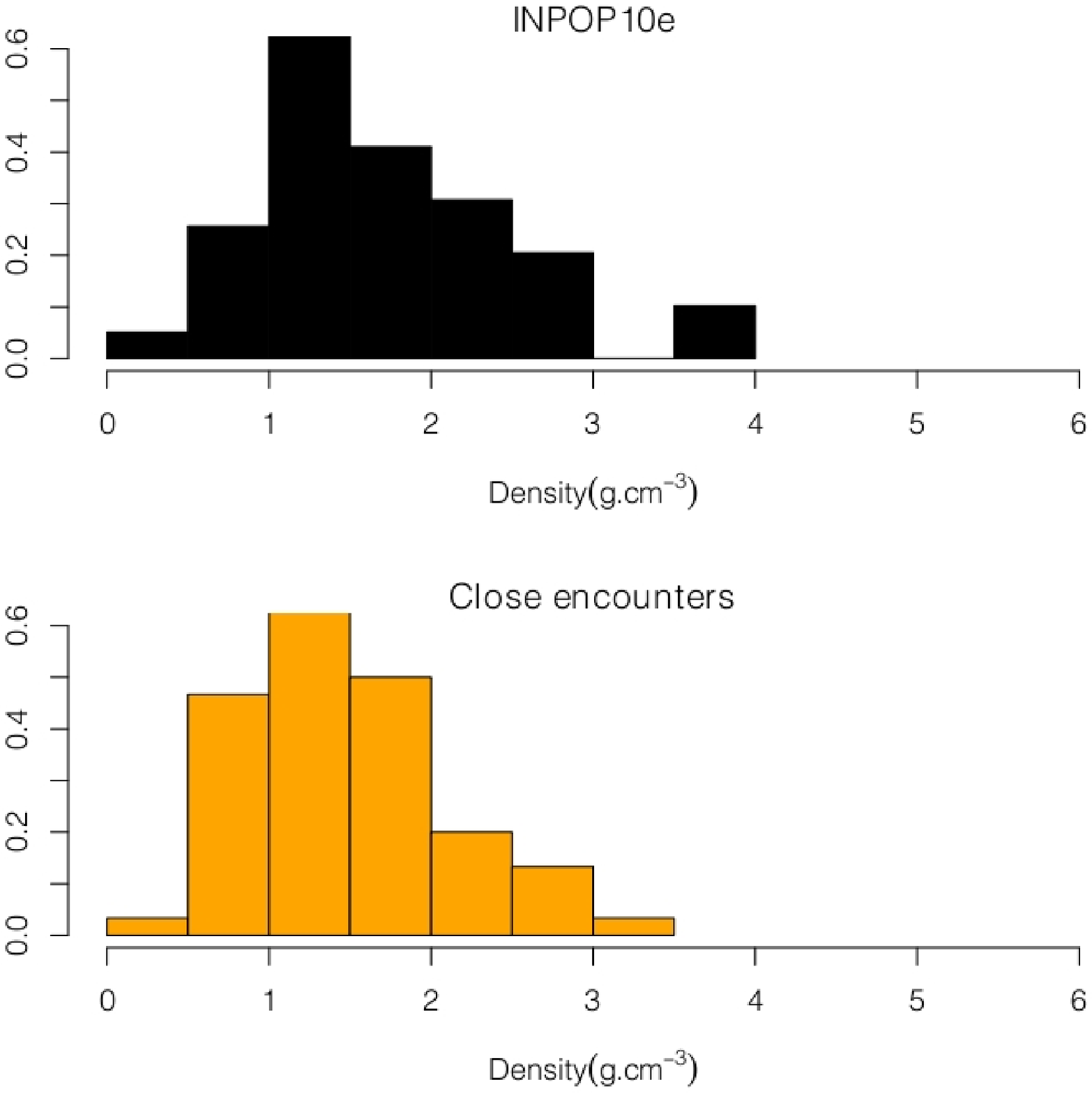}
\caption{INPOP10e Asteroid mass determination: a) INPOP10e asteroid densities compared to other published values versus the impact of the asteroids over the Mars-earth distances during a 1970 to 2012 period. b) Histograms of distribution of the asteroid densities obtained with INPOP10e and with close-encounter methods (right-hand side chart).}
\label{ast}
\end{center}

\end{figure}

Due to the perturbations of the main belt asteroids over the Mars and the Earth orbits, asteroid mass determinations deduced from the construction of planetary ephemerides and the analysis of the high accurate Mars orbiter tracking distances are done regularly (Konopliv et al. 2011, Fienga et al. 2011, Somenzi et al. 2010, Konopliv et al. 2006). However, the inversion problem is here very complex as only less than 50 asteroid masses (to be compared with the 300 asteroids included in the dynamic model of the ephemeris) are known and as all the perturbations cumulate over the Mars geocentric distances. Sophisticated procedures have been tested for years (Standish and Fienga 2002, Kuchynka 2010, Kuchynka et al. 2010). Thanks to the implementation of bounded value least squares associated with a-priori sigma estimators and to the corrections of solar plasma perturbations, we have been able to estimate 152 asteroid masses presented here. This release is quite satisfactory: comparisons between INPOP10e values and values obtained by other authors either by planetary ephemeris construction (Konopliv et al. 2011, Kuchynka 2012) or by close-encounter methods (Zielenbach 2011, Baer et al. 2011) are indeed in good agreement. As one can see on figure \ref{ast}, asteroids inducing more than 7 meter perturbations over the Mars-earth distances have very consistent densities. This limit is consistent with the dispersion of the postfit residuals presented in table \ref{omctab0} which is about 9 meters for all the data sample and 4 meters out of the conjunction periods. Furthermore, for small perturbers, and contrarily to the previous INPOP versions, INPOP10e does not provide unrealistic densities, smaller than 0.5 g.cm$^{-3}$ or greater than 6 g.cm$^{-3}$. A list of the 68 asteroid masses (inducing perturbations bigger than 3 meters over the earth-Mars distances during the 1970 to 2012 observational period) obtained with INPOP10e is given in the appendix \ref{astmass}.

\begin{landscape}

\begin{table}
\caption{Statistics of the residuals obtained after the INPOP10e fit. For comparison, means and standard deviations of residuals obtained with INPOP10a and DE423 are given. In italic are indicated INPOP06b residuals not included in the original fit and which can be seen as INPOP06b extrapolated differences.}
\begin{center}
\begin{tabular}{l c l c | c c | c c | c c} 
\hline
Type of data & & Nbr & Time Interval & \multicolumn{2}{c}{INPOP10e}& \multicolumn{2}{c}{INPOP06b} & \multicolumn{2}{c}{DE423} \\
\hline
Mercury & range [m]& 462 & 1971.29 - 1997.60 & -45.3 & 872.499 & 218.487 & 869.989 & -117.323 & 879.778   \\
Mercury  Mariner & range [m]& 2 & 1974.24 - 1976.21 & -52.486 & 113.185 & -1312.066 & 207.971 & -86.416 & 52.073   \\
Mercury  flybys  Mess & ra [mas]& 3 & 2008.03 - 2009.74 & 0.738 & 1.485 & -0.537 & 0.209 & 0.170 & 1.167   \\
Mercury  flybys  Mess & de [mas]& 3 & 2008.03 - 2009.74 & 2.422 & 2.517 & 1.913 & 2.533 & 1.565 & 2.429   \\
Mercury  flybys  Mess & range [m]& 3 & 2008.03 - 2009.74 & -5.047 & 5.792 & 231.006 & 1466.908 & 22.0 & 14.8   \\
\hline
Venus & VLBI [mas]& 46 & 1990.70 - 2010.86 & 1.590 & 2.602 & -0.634 & 2.834 & 2.166 & 2.518   \\
Venus & range [m]& 489 & 1965.96 - 1990.07 & 500.195 & 2234.924 & 2498.169 & 3671.999 & 496.861 & 2236.798   \\
Venus  Vex & range [m]& 22145 & 2006.32 - 2009.78 & -0.054 & 4.091 & 538.020 & 5246.298 & 1.655 & 4.057   \\
\hline
Mars & VLBI [mas]& 96 & 1989.13 - 2007.97 & -0.004 & 0.407 & -0.408 & 0.535 & -0.319 & 0.457   \\
Mars  Mex & range [m]& 13842 & 2005.17 - 2009.78 & -0.503 & 9.859 & 31.752 & 22.173 & 0.945 & 9.611   \\
Mars  MGS & range [m]& 13091 & 1999.31 - 2006.83 & -0.341 & 3.926 & 17.171 & 16.583 & 0.746 & 4.052   \\
Mars  Ody & range [m]& 5664 & 2006.95 - 2010.00 & 0.280 & 4.155 & 33.399 & 16.995 & 2.021 & 3.504   \\
Mars  Path & range [m]& 90 & 1997.51 - 1997.73 & -6.289 & 13.663 & 9.374 & 13.648 & 23.393 & 13.821   \\
Mars  Vkg & range [m]& 1257 & 1976.55 - 1982.87 & -1.391 & 39.724 & -1.181 & 38.557 & -26.153 & 38.993   \\
\hline
Jupiter & VLBI [mas]& 24 & 1996.54 - 1997.94 & -0.291 & 11.068 & -2.815 & 11.247 & -0.069 & 10.958   \\
Jupiter & ra [arcsec]& 6532 & 1914.54 - 2008.49 & -0.039 & 0.297 & -0.044 & 0.296 & -0.039 & 0.297   \\
Jupiter & de [arcsec]& 6394 & 1914.54 - 2008.49 & -0.048 & 0.301 & -0.045 & 0.302 & -0.048 & 0.301   \\
Jupiter  flybys & ra [mas]& 5 & 1974.92 - 2001.00 & 2.368 & 3.171 & -3.834 & 17.955 & 1.919 & 3.529   \\
Jupiter  flybys & de [mas]& 5 & 1974.92 - 2001.00 & -10.825 & 11.497 & -10.585 & 16.807 & -11.117 & 11.706   \\
Jupiter  flybys & range [m]& 5 & 1974.92 - 2001.00 & -907.0 & 1646.210 & 37467.054 & 55467.239 & -998.461 & 1556.568   \\
\hline
Saturne & ra [arcsec]& 7971 & 1913.87 - 2008.34 & -0.006 & 0.293 & 0.022 & 0.286 & -0.006 & 0.293   \\
Saturne & de [arcsec]& 7945 & 1913.87 - 2008.34 & -0.012 & 0.266 & -0.016 & 0.265 & -0.012 & 0.266   \\
Saturne  VLBI  Cass & ra [mas]& 10 & 2004.69 - 2009.31 & 0.215 & 0.637 & 17.299 & 12.561 & -0.193 & 0.664   \\
Saturne  VLBI  Cass & de [mas]& 10 & 2004.69 - 2009.31 & 0.280 & 0.331 & 8.206 & 7.798 & 0.308 & 0.330   \\
Saturne  Cassini & ra [mas]& 31 & 2004.50 - 2007.00 & 0.790 & 3.879 & 33.123 & 8.024 & 0.314 & 3.876   \\
Saturne  Cassini & de [mas]& 31 & 2004.50 - 2007.00 & 6.472 & 7.258 & 6.437 & 7.731 & 6.329 & 7.283   \\
Saturne  Cassini & range [m]& 31 & 2004.50 - 2007.00 & -0.013 & 18.844 & 214542.236 & 68780.508 & 12.277 & 27.375   \\
\hline
\end{tabular}
\end{center}
\label{omctab0}
\end{table}

\begin{table}
\caption{Statistics of the residuals obtained after the INPOP10e fit. For comparison, means and standard deviations of residuals obtained with INPOP10a and DE423 are given. In italic are indicated INPOP06b residuals not included in the original fit and which can be seen as INPOP06b extrapolated differences. The last four lines indicated as {\em{extrap}} in the first column give residuals obtained by comparisons with observations not used in the fit of the three ephemerides and computed positions. These results illustrate how the accuracy of the ephemerides can be extrapolated out from the fitting interval.}
\begin{center}
\begin{tabular}{l c l c | c c | c c | c c} 
\hline
Type of data & & Nbr & Time Interval & \multicolumn{2}{c}{INPOP10e}& \multicolumn{2}{c}{INPOP06b} & \multicolumn{2}{c}{DE423} \\
\hline
Uranus & ra [arcsec]& 13016 & 1914.52 - 2011.74 & 0.007 & 0.205 & -0.074 & 0.217 & -0.008 & 0.220   \\
Uranus & de [arcsec]& 13008 & 1914.52 - 2011.74 & -0.006 & 0.234 & -0.027 & 0.247 & -0.013 & 0.249   \\
Uranus  flybys & ra [arcsec]& 1 & 1986.07 - 1986.07 & -0.021 & 0.000 & -0.087 & 0.000 & -0.022 & 0.000   \\
Uranus  flybys & de [arcsec]& 1 & 1986.07 - 1986.07 & -0.028 & 0.000 & -0.035 & 0.000 & -0.055 & 0.000   \\
Uranus  flybys & range [m]& 1 & 1986.07 - 1986.07 & 19.738 & 0.000 & 1196925.516 & 0.000 & 22.014 & 0.000   \\
\hline
Neptune & ra [arcsec]& 5395 & 1913.99 - 2007.88 & 0.000 & 0.258 & -0.013 & 0.261 & 0.020 & 0.255   \\
Neptune & de [arcsec]& 5375 & 1913.99 - 2007.88 & -0.000 & 0.299 & -0.028 & 0.303 & -0.010 & 0.306   \\
Neptune  flybys & ra [arcsec]& 1 & 1989.65 - 1989.65 & -0.012 & 0.000 & -0.091 & 0.000 & -0.010 & 0.000   \\
Neptune  flybys & de [arcsec]& 1 & 1989.65 - 1989.65 & -0.005 & 0.000 & -0.044 & 0.000 & -0.018 & 0.000   \\
Neptune  flybys & range [m]& 1 & 1989.65 - 1989.65 & 69.582 & 0.000 & -2333073.041 & 0.000 & -121.987 & 0.000   \\
\hline
Pluto & ra [arcsec]& 2458 & 1914.06 - 2008.49 & 0.034 & 0.654 & 0.005 & 0.601 & 0.072 & 0.609   \\
Pluto & de [arcsec]& 2462 & 1914.06 - 2008.49 & 0.007 & 0.539 & -0.024 & 0.519 & -0.011 & 0.521   \\
Pluto  Occ & ra [arcsec]& 13 & 2005.44 - 2009.64 & 0.003 & 0.047 & -0.052 & 0.045 & -0.054 & 0.044   \\
Pluto  Occ & de [arcsec]& 13 & 2005.44 - 2009.64 & -0.006 & 0.018 & 0.032 & 0.032 & 0.006 & 0.028   \\
Pluto  HST & ra [arcsec]& 5 & 1998.19 - 1998.20 & -0.033 & 0.043 & -0.042 & 0.044 & -0.030 & 0.043   \\
Pluto  HST & de [arcsec]& 5 & 1998.19 - 1998.20 & 0.028 & 0.048 & -0.033 & 0.048 & -0.028 & 0.048   \\
\hline
\hline
Venus  Vex & range [m]& 2825 & 2009.78 - 2011.45 & 7.605 & 32.821 & -526.584 & 5846.331 & 12.488 & 32.680   \\
Mars  Mex & range [m]& 57229 & 2009.78 - 2012.43 & -2.95 & 30.14 & 29.581 & 42.313 & 1.508 & 30.902   \\
\hline
\end{tabular}
\end{center}
\label{omctab1}
\end{table}
\end{landscape}

\section{Conclusions}

Planetary ephemerides are not a only crucial tool for celestial mechanics or the preparation of space missions. They can also play an important role in testing gravity, studying the asteroid physics by estimating their masses or in solar physics with the analysis of the solar corona signatures over radiometric tracking observations of planet orbiters. 
We present here the latest INPOP version. It appears to be as accurate as the JPL DE ephemerides and allows several applications in solar physics, planetology and gravity testing. At the end of 2012, the analysis of the Messenger tracking data should be completed and implemented in INPOP. These new observations would be crucial especially for gravity tests. We will also implement the estimation of possible variation of the gravitational mass of the sun. This parameter  would give stringent limits to theoretical developments predicting variations with time of the gravitational constant. More observations of Saturn deduced from the Solstice extended Cassini mission should also be available. These data would be very helpful to maintain the accuracy of the outer planet orbits.
For the Moon, new models of rotation are under development and the use of observations deduced from Moon orbiter LRO radio and laser tracking could be considered as good complements to LLR.

\appendix
\section{Asteroid masses obtained with INPOP10e}
\label{astmass}
\begin{landscape}

\begin{table*}
\caption{Asteroid masses obtained with INPOP10 and and compared with values found in the recent literature. The last column gives the impact of each asteroid on the  Earth-Mars distances over the 1970 to 2010 period. In this table are given only the masses of the asteroids inducing an impact greater than 3 meters. The uncertainties are given at 1 published sigma. The masses presented here are the most significant determinations (with S/N bigger than 1.8) done with INPOP10. Z11 stands for (Zielenbach 2011), B11 for (Baer et al. 2011), K11 for (Konopliv et al. 2011)  and  K12 for (Kuchynka 2012). M08 stands for (Marchis et al 2008) and P11 for (Paetzold et al. 2011)}
\begin{center}
\begin{tabular}{c c c c c c c c c c }
\hline
IAU designation	& INPOP10e	& & Z11	& B11	&	 K11 &	 Others & K12 &  Impact\\
number & $10^{12}$ x M$_{\odot}$	& $\%$ & $10^{12}$ x M$_{\odot}$	 & $10^{12}$ x M$_{\odot}$	& $10^{12}$ x M$_{\odot}$ & $10^{12}$ x M$_{\odot}$ & $10^{12}$ x M$_{\odot}$ & m\\
\hline
4 & 130.274 $\pm$ 0.85 & 0.65 	& 130.270$\pm$   0.71& 130.000$\pm$   0.53& 130.970$\pm$   2.06& 130.27167 $\pm$ 0.0003 \citep{2012Sci...336..684R}& 130.508  $\pm$ 0.82& 1198.95\\
 1 & 467.267 $\pm$ 1.85 & 0.40 	& 473.485$\pm$   1.33& 475.700$\pm$   0.70& 467.900$\pm$   3.25& &473.053 $\pm$ 2.87 &793.74\\
2 & 102.654 $\pm$ 1.60 & 1.56 	& 103.374$\pm$   6.92& 101.000$\pm$   6.50& 103.440$\pm$   2.55& &101.724 $\pm$ 2.11  &146.27\\
324 & 4.769 $\pm$ 0.43 & 9.13 	& 5.422$\pm$   1.00&   &   5.340$\pm$	0.99&  &5.124 $\pm$ 0.38 &93.54\\
10 & 43.997 $\pm$ 3.23 & 7.20 	& 41.286$\pm$	1.47&  43.580$\pm$   0.74&  44.970$\pm$   7.76& &52.821 $\pm$ 4.22   &77.00\\
19 & 4.892 $\pm$ 0.51 & 10.62 	& 5.090$\pm$   0.47&   4.180$\pm$   0.36&   3.200$\pm$   0.53& & 3.918 $\pm$ 0.45 &59.07\\
3 & 11.793 $\pm$ 0.62 & 5.25 	& 15.574$\pm$   1.63&  14.400$\pm$   2.30&  12.100$\pm$   0.91& &13.488 $\pm$ 0.83  &55.64\\
704 & 19.217 $\pm$ 2.37 & 12.32 	 & 15.738$\pm$   2.61&  19.650$\pm$   0.89&  19.970$\pm$   6.57& &19.817 $\pm$ 3.467  &34.49\\
532 & 11.552 $\pm$ 1.18 & 9.21 	 & 8.794$\pm$	2.18&  16.800$\pm$   2.80&   4.970$\pm$   2.81& & 6.481 $\pm$ 1.36 &32.71\\
9 & 4.202 $\pm$ 0.67 & 15.24 	& 4.524$\pm$   0.67&   5.700$\pm$   1.10&   3.280$\pm$	1.08&  &3.467 $\pm$ 0.68 &29.61\\
7 & 6.302 $\pm$ 0.61 & 9.68 	&8.434$\pm$   0.80&   8.120$\pm$   0.46&   5.530$\pm$   1.32& &7.459 $\pm$ 0.83  &27.82\\
29 & 7.227 $\pm$ 1.04 & 14.36 	& 5.552$\pm$   0.82&   7.630$\pm$   0.31&   7.420$\pm$   1.49& &5.199 $\pm$ 1.20 &26.67\\
31 & 13.234 $\pm$ 1.98 & 14.95	 &   13.512$\pm$   4.57&  29.200$\pm$	9.90&  &&11.001 $\pm$ 3.69 &   23.47\\
13 & 4.713 $\pm$ 0.78 & 31.97 	& 3.054 $\pm$ 1.61 & 8.0 $\pm$ 2.2 & & & 6.18 $\pm$ 1.66 & 22.04 \\
15 & 15.839 $\pm$ 0.95 & 6.00 	& 16.178$\pm$	0.40&  15.597$\pm$   0.10&  14.180$\pm$   1.49& &13.111 $\pm$ 1.43  &21.55\\
6 & 7.084 $\pm$ 0.84 & 11.81 	& 7.733$\pm$   1.22&   6.400$\pm$   0.67&   6.730$\pm$	1.64& &4.219 $\pm$ 0.98  &21.15\\
139 & 2.130 $\pm$ 0.88 & 36.52 	 & 3.015  $\pm$ 1.6 & & & & & 16.7\\
747 & 0.723 $\pm$ 0.64 & 64.12 	 & 2.639$\pm$	2.25&	&   &	&&15.94\\
105 & 3.046 $\pm$ 0.64 & 20.42 	 &   &   &   &   &&15.20\\
20 & 2.897 $\pm$ 1.05 & 48.27 	& 3.032 $\pm$ 0.57 & 1.680 $\pm$ 0.35 & & & &14.8  \\
8 & 3.357 $\pm$ 0.39 & 11.72 & 3.693$\pm$   0.66&   3.330$\pm$   0.42&   2.010$\pm$   0.42& &2.185 $\pm$ 0.38 &  12.66\\
405 & 1.378 $\pm$ 0.33 & 21.52 & &   &   &   &   & 11.38\\
511 & 18.250 $\pm$ 2.87 & 15.70 & 13.143$\pm$   3.03&  18.960$\pm$   0.90&   8.580$\pm$   5.93& &14.844 $\pm$ 4.2 & 10.25\\
52 & 10.682 $\pm$ 2.58 & 24.20 & 13.957$\pm$   1.63&  11.390$\pm$   0.79&  11.170$\pm$   8.40& &16.728 $\pm$ 4.1  &  9.84\\
16 & 12.613 $\pm$ 2.20 & 17.41 & 12.279$\pm$   0.81&  11.400$\pm$   0.42&  12.410$\pm$   3.44& &8.891 $\pm$ 2.11 &  9.70\\
\hline
  \end{tabular}
\end{center}
\label{mass0}
\end{table*}
  \end{landscape}

\begin{landscape}

\begin{table*}
\caption{}
\begin{center}
\begin{tabular}{c c c c c c c c c c }
\hline
IAU designation	& INPOP10e	& & Z11	& B11	&	 K11 &	 Others & K12 &  Impact\\
number	& $10^{12}$ x M$_{\odot}$	& $\%$ & $10^{12}$ x M$_{\odot}$	 & $10^{12}$ x M$_{\odot}$	& $10^{12}$ x M$_{\odot}$ & $10^{12}$ x M$_{\odot}$ & $10^{12}$ x M$_{\odot}$ & m\\
\hline
419 & 1.649 $\pm$ 0.44 & 22.92 	 & &   &   &	&&9.59\\
78 & 2.562 $\pm$ 0.57 & 17.75 	& &   &   &    &&9.39\\
23 & 1.545 $\pm$ 0.34 & 20.12 	& &   &   &  &0.829 $\pm$ 0.30  &9.07\\
488 & 5.157 $\pm$ 1.81 & 29.04 	 & 1.099 $\pm$ 0.34 &	&   &  &  &8.61 \\
409 & 0.001 $\pm$ 0.001 & 49.38 	 & 6.211 $\pm$ 1.63 &	&   &  &  & 7.57\\
94 & 15.032 $\pm$ 3.44 & 30.80 	 & &   &   & &   &7.47\\
111 & 4.489 $\pm$ 1.18 & 26.27 	 & & & & & & 6.98 \\
109 & 0.161 $\pm$ 0.25 & 135.81 	 & & & & & & 6.86\\
63 & 0.003 $\pm$ 0.00 & 61.13 	& &   &   &  &  &6.45\\
12 & 0.524 $\pm$ 0.30 & 62.86 	& 0.825$\pm$   1.18&   &   & &  & 6.16\\
469 & 0.004 $\pm$ 0.00 & 46.73 	 & & & & & & 6.11\\
356 & 4.173 $\pm$ 0.58 & 14.03 	 &&   &   &  &  &5.76\\
88 & 7.088 $\pm$ 1.42 & 22.27 	& 5.65 $\pm$ 1.66 &&&&&5.74\\
128 & 6.859 $\pm$ 1.58 & 29.02 	 & 4.213 $\pm$ 1.078 & & & && 5.63\\
194 & 5.601 $\pm$ 0.64 & 10.66 	& &   &   &   &    &5.14\\
51 & 0.009 $\pm$ 0.00 & 46.49 	&   1.687$\pm$   0.81&   &   &    &&5.11\\
156 & 3.263 $\pm$ 0.44 & 13.60 	 & &   &   &   &   & 5.10\\
516 & 0.350 $\pm$ 0.14 & 20.20 	 & & & & & & 5.05\\
451 & 14.984 $\pm$ 3.60 & 24.03 	 & 5.604$\pm$   3.22&  10.200$\pm$   3.40&   &  & & 4.74\\
313 & 1.022 $\pm$ 0.89 & 124.30 	 & & & & & & 4.70\\
107 & 3.413 $\pm$ 1.51 & 44.19 	 & 8.846$\pm$	4.37&	&   & 5.630$\pm$   0.10 (M08) & &  4.63\\
65 & 4.210 $\pm$ 0.86 & 16.29 	& 7.652$\pm$   1.73&   5.300$\pm$   0.96&   &  & & 4.54\\
21 & 0.867 $\pm$ 0.79 & 91.18 	& & 1.31 $\pm$ 0.44 && 0.8547 $\pm$ 0.0085 (P11) & & 4.53\\
694 & 0.000 $\pm$ 0.00 & 156.17 	 & & & & & & 4.19\\
134 & 1.014 $\pm$ 0.37 & 29.71 	 &&   &   &   && 4.17\\
54 & 8.392 $\pm$ 1.08 & 12.91 	& 1.480$\pm$   1.58&   &   &  &  &4.09\\
\hline
  \end{tabular}
\end{center}
\label{mass1}
\end{table*}

  \end{landscape}
  
  \begin{landscape}

\begin{table*}
\caption{ }
\begin{center}
\begin{tabular}{c c c c c c c c c c }
\hline
IAU designation	& INPOP10e	& & Z11	& B11	&	 K11 &	 Others & K12 &  Impact\\
number	& $10^{12}$ x M$_{\odot}$	& $\%$ & $10^{12}$ x M$_{\odot}$	 & $10^{12}$ x M$_{\odot}$	& $10^{12}$ x M$_{\odot}$ & $10^{12}$ x M$_{\odot}$ & $10^{12}$ x M$_{\odot}$ & m\\
\hline
106 & 3.87 $\pm$ 0.41 & 29.90 	 & 1.769 $\pm$ 1.319 & & & & & 3.90 \\
173 & 6.743 $\pm$ 1.50 & 22.32 	 & 0.669$\pm$	0.61&	&   &	& &3.81\\
22 & 8.374 $\pm$ 0.65 & 9.04 	& 6.592$\pm$   1.93&    &  & 4.07$\pm$	0.1 (M08) &  &3.76\\
444 & 5.329 $\pm$ 1.84 & 28.26 	 & 5.608$\pm$	1.34&	&   &  &  &3.73\\
185 & 4.407 $\pm$ 1.26 & 22.55 	 & 3.463 $\pm$ 3.1& & & & &3.65\\
46 & 3.525 $\pm$ 0.74 & 21.07 	&&   &   &  &  &3.56\\
37 & 2.343 $\pm$ 0.60 & 26.36 	 & &   &   &  & & 3.55\\
53 & 2.007 $\pm$ 0.90 & 48.25 	 & & & & & & 3.55\\
164 & 0.001 $\pm$ 0.00 & 65.12 	 && & & & & 3.43\\
410 & 3.476 $\pm$ 0.66 & 19.05 	 & &   &   &  &  &3.39\\
85 & 2.190 $\pm$ 0.81 & 45.02 	 & 1.755$\pm$	1.33&	&   &  &  &3.38\\
1021 & 0.551 $\pm$ 0.94 & 197.24 	& &   &   &  & & 3.25\\
56 & 2.676 $\pm$ 0.61 & 22.86 	& & & & & & 3.22\\
34 & 1.452 $\pm$ 0.61 & 59.17 	& &   &   &  & & 3.15\\
17 & 3.686 $\pm$ 0.92 & 24.98 	& & & & & & 3.03\\
404 & 0.628 $\pm$ 0.59 & 130.64 	 & &   &  &  &&3.01\\
200 & 0.574 $\pm$ 0.07 & 12.98 	& &   &  & && 3.00\\
\hline
  \end{tabular}
\end{center}
\label{mass2}
\end{table*}
  \end{landscape}

\begin{table}
\caption{Values of dynamical parameters fitted to LLR observations. $GM_{EMB}$ is the sum of Earth's and Moon's masses, multiplied by the gravitationnal constant and is expressed in AU\textsuperscript{3}/day\textsuperscript{2}.
$C_{nmE}$ are the Earth's coefficients of potential (without unit).
$\tau_{21E}$ and $\tau_{22E}$ are time delays of the Earth used for tides effects and expressed in days.
$C_{nmM}$ and $S_{nmM}$ are the Moon's coefficients of potential (without unit). $(C/MR^2)_M$ is the ratio between the third moment of inertia of the Moon, divided by its mass and the square of the mean equatorial radius (without unit).
$k_{2M}$ and $\tau_M$ are the Love number (without unit) and the time delay (in day) of the Moon.
Formal errors at $1\sigma$ are given if the parameter is fitted and correspond to the values provided by the covariance matrix of the least square fit;
one can note that the real uncertainties on parameters are generally much higher.
Fixed values come from Lunar gravity model LP150Q (Konopliv et al. 2001) and Earth's ones from EGM96 ($\textrm{cddis.nasa.gov/926/egm96}$).}
\begin{center}
\begin{tabular}{c|r|l}
Name & \multicolumn{1}{c|}{Value} & \multicolumn{1}{c}{Formal error ($1\sigma$)} \\
\hline
$ GM_{EMB} $ 	 & $  8.9970115965 \times 10^{-10} $ & $ \pm 7.3 \times 10^{-19} $ \\ 
$ C_{20E} $ 	 & $ -1.0826222 \times 10^{-3} $ & $ \pm 1.0 \times 10^{-9} $ \\ 
$ C_{30E} $ 	 & $  2.875 \times 10^{-6} $ & $ \pm 3.9 \times 10^{-8} $ \\ 
$ C_{40E} $ 	 & $  1.6196215913670001 \times 10^{-6} $ &   \\ 
$ \tau_{21E} $ 	 & $  1.1841 \times 10^{-2} $ & $ \pm 8.8 \times 10^{-5} $ \\ 
$ \tau_{22E} $ 	 & $  7.0163 \times 10^{-3} $ & $ \pm 7.2 \times 10^{-6} $ \\ 
$ C_{20M} $ 	 & $ -2.03443 \times 10^{-4} $ & $ \pm 2.7 \times 10^{-8} $ \\ 
$ C_{22M} $ 	 & $  2.23971 \times 10^{-5} $ & $ \pm 2.6 \times 10^{-9} $ \\ 
$ C_{30M} $ 	 & $ -8.396 \times 10^{-6} $ & $ \pm 2.3 \times 10^{-8} $ \\ 
$ C_{31M} $ 	 & $  3.191 \times 10^{-5} $ & $ \pm 3.7 \times 10^{-7} $ \\ 
$ C_{32M} $ 	 & $  4.8452131769807101 \times 10^{-6} $ &   \\ 
$ C_{33M} $ 	 & $  1.7279 \times 10^{-6} $ & $ \pm 6.2 \times 10^{-9} $ \\ 
$ C_{40M} $ 	 & $  9.6422863508400007 \times 10^{-6} $ &   \\ 
$ C_{41M} $ 	 & $ -5.6926874002713197 \times 10^{-6} $ &   \\ 
$ C_{42M} $ 	 & $ -1.5861997682583101 \times 10^{-6} $ &   \\ 
$ C_{43M} $ 	 & $ -8.1204110561427604 \times 10^{-8} $ &   \\ 
$ C_{44M} $ 	 & $ -1.2739414703200301 \times 10^{-7} $ &   \\ 
$ S_{31M} $ 	 & $  3.167 \times 10^{-6} $ & $ \pm 8.5 \times 10^{-8} $ \\ 
$ S_{32M} $ 	 & $  1.68722 \times 10^{-6} $ & $ \pm 5.7 \times 10^{-10} $ \\ 
$ S_{33M} $ 	 & $ -2.4855254931699199 \times 10^{-7} $ &   \\ 
$ S_{41M} $ 	 & $  1.5743934836970999 \times 10^{-6} $ &   \\ 
$ S_{42M} $ 	 & $ -1.5173124037059000 \times 10^{-6} $ &   \\ 
$ S_{43M} $ 	 & $ -8.0279066452763596 \times 10^{-7} $ &   \\ 
$ S_{44M} $ 	 & $  8.3147478750240001 \times 10^{-8} $ &   \\ 
$ (C/MR^2)_{M} $ & $  3.93129 \times 10^{-1} $ & $ \pm 4.6 \times 10^{-5} $ \\ 
$ k_{2M} $ 	 & $  2.656 \times 10^{-2} $ & $ \pm 1.7 \times 10^{-4} $ \\ 
$ \tau_{M} $ 	 & $  1.881 \times 10^{-1} $ & $ \pm 1.2 \times 10^{-3} $ \\ 
\hline
\end{tabular}
\end{center}
\label{Tab_valeurs_parametres_LLR_dyn_I10e}
\end{table}

\begin{table}
\caption{Selenocentric coordinates of reflectors, expressed in meters.}
\begin{center}
\begin{tabular}{cc|r|l}
Reflector & &\multicolumn{1}{c|}{Value} & \multicolumn{1}{c}{Formal error ($1\sigma$)} \\
\hline
           & x & $  1591924.289 $ & $ \pm 1.450 $ \\ 
Apollo XI  & y & $   690804.320 $ & $ \pm 3.330 $ \\ 
           & z & $    21002.773 $ & $ \pm 0.303 $ \\ 
\hline
           & x & $  1652726.748 $ & $ \pm 1.100 $ \\ 
Apollo XIV & y & $  -520888.942 $ & $ \pm 3.460 $ \\ 
           & z & $  -109731.546 $ & $ \pm 0.314 $ \\ 
\hline
           & x & $  1554675.389 $ & $ \pm 0.269 $ \\ 
Apollo XV  & y & $    98197.798 $ & $ \pm 3.250 $ \\ 
           & z & $   765004.968 $ & $ \pm 0.302 $ \\ 
\hline
           & x & $  1114347.290 $ & $ \pm 1.650 $ \\ 
Lunakhod 1 & y & $  -781225.859 $ & $ \pm 2.330 $ \\ 
           & z & $  1076058.870 $ & $ \pm 0.254 $ \\ 
\hline
           & x & $  1339314.213 $ & $ \pm 1.690 $ \\ 
Lunakhod 2 & y & $   801960.228 $ & $ \pm 2.800 $ \\ 
           & z & $   756358.542 $ & $ \pm 0.265 $ \\ 
\hline
\end{tabular}
\end{center}
\label{Tab_valeurs_parametres_LLR_I10e_reflecteurs}
\end{table}

\begin{table}
\caption{ITRF coordinates of stations at J1997.0, expressed in meters.}
\begin{center}
\begin{tabular}{cc|r|l}
Station & &\multicolumn{1}{c|}{Value} & \multicolumn{1}{c}{Formal error ($1\sigma$)} \\
\hline
                  & x & $  4581692.117 $ & $ \pm 0.003 $ \\ 
Cerga             & y & $   556196.022 $ & $ \pm 0.001 $ \\ 
                  & z & $  4389355.034 $ & $ \pm 0.010 $ \\ 
\hline
                  & x & $ -1330781.431 $ & $ \pm 0.011 $ \\ 
Mc Donald         & y & $ -5328755.466 $ & $ \pm 0.009 $ \\ 
                  & z & $  3235697.521 $ & $ \pm 0.022 $ \\ 
\hline
                  & x & $ -1330121.100 $ & $ \pm 0.014 $ \\ 
MLRS1             & y & $ -5328532.269 $ & $ \pm 0.008 $ \\ 
                  & z & $  3236146.586 $ & $ \pm 0.024 $ \\ 
\hline
                  & x & $ -1330021.433 $ & $ \pm 0.002 $ \\ 
MLRS2             & y & $ -5328403.284 $ & $ \pm 0.003 $ \\ 
                  & z & $  3236481.626 $ & $ \pm 0.010 $ \\ 
\hline
                  & x & $ -5466000.427 $ & $ \pm 0.011 $ \\ 
Haleakala (rec.)  & y & $ -2404424.705 $ & $ \pm 0.013 $ \\ 
                  & z & $  2242206.715 $ & $ \pm 0.028 $ \\ 
\hline
                  & x & $ -1463998.834 $ & $ \pm 0.004 $ \\ 
Apollo            & y & $ -5166632.673 $ & $ \pm 0.004 $ \\ 
                  & z & $  3435013.105 $ & $ \pm 0.011 $ \\ 
\hline
                  & x & $  4641979.025 $ & $ \pm 0.093 $ \\ 
Matera            & y & $  1393067.140 $ & $ \pm 0.250 $ \\ 
                  & z & $  4133250.084 $ & $ \pm 0.229 $ \\ 
\hline
\end{tabular}
\end{center}
\label{Tab_valeurs_parametres_LLR_I10e_stations}
\end{table}

\begin{table}
\caption{Means and standard deviations (both expressed in centimeters) of LLR residuals for INPOP10e solution. Na is the total number of observations available, Nk is the number kept in fitting process, Nr is the number that have been rejected according to the $3\sigma$ criterion (Na is always Nk+Nr).}
\begin{center}
\begin{tabular}{c|c|r|r|r|r|r}
Station & Period & Mean & Std. dev. & Na & Nk & Nr \\
\hline
Cerga      &  1987-1995  &  -0.45 &   6.35 & 3460 & 3415 &   45 \\
Cerga      &  1995-2012  &   0.05 &   4.01 & 5143 & 5058 &   85 \\
Cerga      &  1984-1986  &   7.10 &  15.89 & 1187 & 1158 &   28 \\
Mc Donald  &  1969-1986  &   0.20 &  31.25 & 3604 & 3487 &  117 \\
MLRS1      &  1982-1985  &  -7.10 &  73.41 &  418 &  405 &   13 \\
MLRS1      &  1985-1988  &   0.24 &   7.35 &  174 &  163 &   11 \\
MLRS2      &  1988-1996  &  -0.41 &   4.71 & 1192 & 1148 &   44 \\
MLRS2      &  1996-2012  &   0.18 &   5.58 & 2498 & 1972 &  526 \\
Haleakala  &  1984-1990  &  -0.40 &   8.09 &  770 &  733 &   37 \\
Apollo     &  2006-2010  &   0.07 &   5.22 &  942 &  935 &    7 \\
Matera     &  2003-2012  &  -0.30 &  29.50 &   33 &   26 &    7 \\
\end{tabular}
\end{center}
\label{Tab_residus_LLR_I10e}
\end{table}

\begin{figure}
\begin{center}
\includegraphics[scale=0.9]{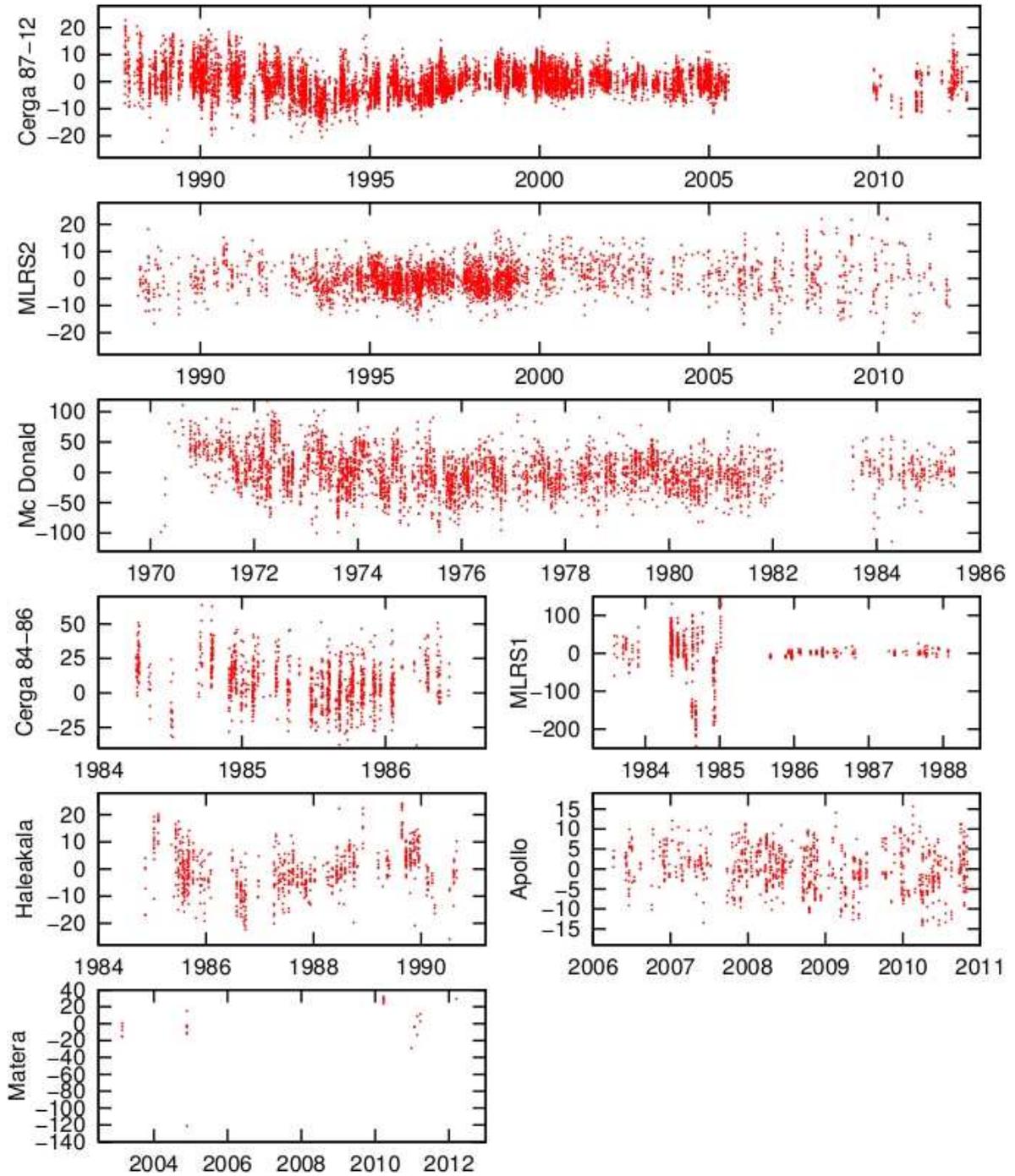}
\caption{Postfit LLR resdiduals with INPOP10e for each station, expressed in centimeters.}
\label{FigResidusLLR10e}
\end{center}
\end{figure}

\begin{figure}
\begin{center}
\includegraphics[scale=1.5]{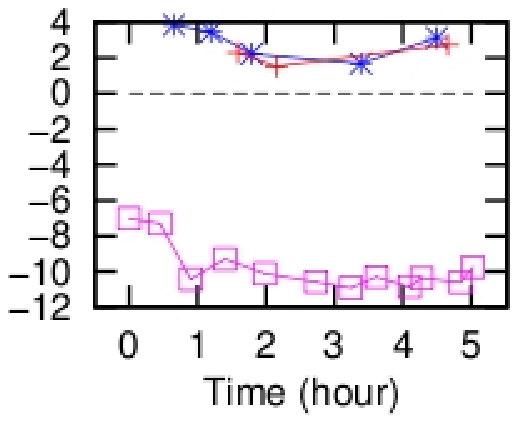}\includegraphics[scale=1.5]{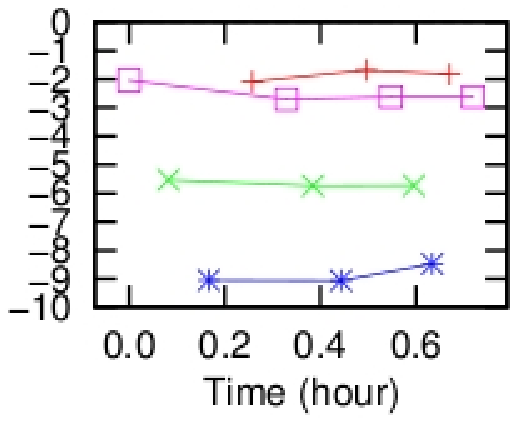}
\caption{LLR residuals (in centimeters) for two different nights: Cerga on 11 february 2011 on the left, Apollo on 18 october 2010 on the right. The duration of the  observation session is 5 hours for Cerga, 0.7 hour for Apollo. Each line corresponds to a different reflector: Apollo XI (red), Lunakhod 1 (green), Apollo XIV (blue) and Apollo XV (megenta).}
\label{FigResidusLLR10e2Nuits}
\end{center}
\end{figure}

\end{document}